\documentclass[aps,twocolumn,superscriptaddress,floatfix,prl]{revtex4-1}

\usepackage{lipsum}
\usepackage{graphicx}
\usepackage{amsmath,amssymb}
\usepackage{braket}
\usepackage{mathtools}
\usepackage{hyperref}
\usepackage{multirow}
\usepackage{color}
\usepackage[normalem]{ulem}
\usepackage{amsfonts}
\usepackage{float}
\usepackage{pdfpages} % include appendix
\makeatletter
\usepackage{subfigure}

\def\beq{\begin{equation}}
\def\eeq{\end{equation}}
\def\bea{\begin{eqnarray}}
\def\eea{\end{eqnarray}}

\AtBeginDocument{\let\LS@rot\@undefined}
\makeatother

\begin{document}

%\title{Generalised hydrodynamics of spin helices}

\title{Hydrodynamic relaxation of spin helices}

\author{Guillaume Cecile}
\affiliation{Laboratoire de Physique Th\'eorique et Mod\'elisation, CNRS UMR 8089,CY Cergy Paris
Universit\'e, 95302 Cergy-Pontoise Cedex, France}

\author{Sarang Gopalakrishnan}
%\affiliation{Department of Physics, The Pennsylvania State University, University Park, PA 16802, USA}
\affiliation{Department of Electrical and Computer Engineering, Princeton University, Princeton, NJ 08544, USA}

\author{Romain Vasseur}
\affiliation{Department of Physics, University of Massachusetts, Amherst, MA 01003, USA}

\author{Jacopo De Nardis}
\affiliation{Laboratoire de Physique Th\'eorique et Mod\'elisation, CNRS UMR 8089,CY Cergy Paris
Universit\'e, 95302 Cergy-Pontoise Cedex, France}

\begin{abstract}

Motivated by recent cold atom experiments, we study the relaxation of spin helices in quantum XXZ spin chains. 
The experimentally observed relaxation of spin helices follows scaling laws that are qualitatively different from linear-response transport.
We construct a theory of the relaxation of helices, combining generalized hydrodynamics (GHD) with diffusive corrections and the local density approximation. 
Although helices are far from local equilibrium (so GHD need not apply \emph{a priori}), our theory reproduces the experimentally observed relaxational dynamics of helices. 
In particular, our theory explains the existence of temporal regimes with apparent anomalous diffusion, as well as the asymmetry between positive and negative anisotropy regimes. 

\end{abstract}
\vspace{1cm}

\maketitle
\paragraph{\textbf{Introduction} ---}
How interacting, isolated quantum systems relax from far-from-equilibrium initial states is one of the basic problems in many-body physics. This problem is particularly interesting in one dimension, since many experimentally relevant one-dimensional systems are approximately integrable. 
%
%The study of the importance of quantum effects on this relaxation process was reinvigorated by the experimental realizations of ultracold atomic gases, including cases where the atoms are confined to one dimension where the dynamics is controlled by a Hamiltonian close to integrability.
Integrable systems support stable, ballistically propagating quasiparticles even at high energy density. The presence of ballistic quasiparticles might suggest that transport of the conserved charges should be ballistic; however, because of dressing effects due to interactions this is not always the case. In many systems, such as anisotropic XXZ spin chains, spin transport can be either ballistic, diffusive or even superdiffusive depending on the anisotropy. The nature of finite-temperature spin transport in the XXZ spin chain has been studied extensively very recently,  both experimentally~\cite{bloch_heisenberg,Jepsen2020,
Wei2022,Scheie2021,Jepsen2022,2022arXiv220909322P} and theoretically \cite{PhysRevX.11.031023,PhysRevLett.107.250602,Ljubotina2019,PhysRevB.104.115163,PhysRevB.103.235115,PhysRevB.57.8307,PhysRevB.102.115121,bulch2019superdiffusive,Bulchandani2021,RevModPhys.93.025003,PhysRevB.104.L081410,PhysRevB.104.195409}. The theoretical analysis of transport in nearly integrable systems relies on generalized hydrodynamics (GHD)~\cite{PhysRevX.6.041065,PhysRevLett.117.207201,Bastianello_2022}, a description of the asymptotic late-time dynamics that is expected to apply once the system has locally approached a generalized Gibbs ensemble (GGE)~\cite{VR_review}.

In cold-atom experiments, it is often more convenient to prepare a pure initial state than a thermal (mixed) one. 
A class of states that can straightforwardly be prepared are spin helices in which the spin orientation varies spatially in a periodic manner with a given wavelength $\lambda$~\cite{bloch_heisenberg}. 
These states are far from local thermal equilibrium, so it is not \emph{a priori} obvious that GHD can describe their dynamics. 
On general grounds, we expect such helices to relax with a rate $\Gamma \sim \lambda^{-z}$ with $z$ a dynamical exponent which need not coincide with the linear-response one since the spiral is very far from equilibrium. Recent experimental results \cite{Jepsen2020} indicate a particularly rich behavior as a function of the anisotropy, different from linear-response GHD expectations, which so far eluded any theoretical explanation. 

The  XXZ spin-$\frac{1}{2}$ chain is described by the Hamiltonian (in the following we shall set $J=1$)
\beq\label{eqXXZ}
H =  J \sum_i \Big[ S^x_i S^x_{i+1} + S^y_i S^y_{i+1} + \Delta S^z_i S^z_{i+1}\Big].
\eeq
Energy transport in the XXZ spin chain is purely ballistic regardless of $\Delta$, as the energy current is conserved under the dynamics. Spin transport at half-filling and high-temperature, however, depends much more nontrivially on the anisotropy $\left| \Delta \right|$ (but not on its sign), see~\cite{Bertini2021,2022arXiv220811133G} for recent reviews. In the easy-plane regime $\left| \Delta \right| < 1$, spin transport has a ballistic component~\cite{PhysRevLett.111.057203,IN_Drude}, while for $\left| \Delta \right| >1$ it is believed to be diffusive~\cite{DeNardis_SciPost,GV19}. The isotropic point $\left| \Delta \right| =1$ corresponds to a dynamical phase transition characterized by superdiffusive transport with dynamical exponent $z=3/2$~\cite{Ljubotina_nature,Ljubotina19,GV19,Wei2022,Scheie2021,Bulchandani2021}. Experimental results pertaining to spin helices relaxation reveal a very different picture~\cite{bloch_heisenberg,Jepsen2020}: Spin helices appear to relax (1)  diffusively $\Gamma \sim \lambda^{-2}$ at the antiferromagnetic isotropic point $\Delta=1$~\cite{bloch_heisenberg}, (2) subdiffusively ($z>2$) at short times for $\Delta>1$, (3) superdiffusively with $1<z<2$ for anisotropy $0<\Delta<1$, and (4) ballistically $\Gamma \sim \lambda^{-1}$ for $\Delta \leq 0$, with a crossover to diffusive relaxation at longer times for $\Delta < -1$. The nature of the relaxation appears to be in sharp contrast with linear response (GHD) results, suggesting that a different mechanism, possibly beyond hydrodynamics and/or of quantum mechanical nature is at play. 

In this letter, we present a theory of how spin helices relax, using a local-density approximation (LDA) assuming local equilibration and GHD equations including diffusive corrections. We assume that spin helices first relax to local equilibrium, and can be described by a local GGE with spatially varying Lagrange multipliers. The remaining time evolution is obtained from GHD by numerical integration.  
Our theory reproduces all the qualitative features observed in ultracold atom experiments, and sheds light on the asymptotic long-time behavior. In particular, our results reconcile the apparent contradiction between GHD and the recent experiments~\cite{bloch_heisenberg,Jepsen2020}.

\begin{figure*}[t]
  \includegraphics[width=\textwidth]{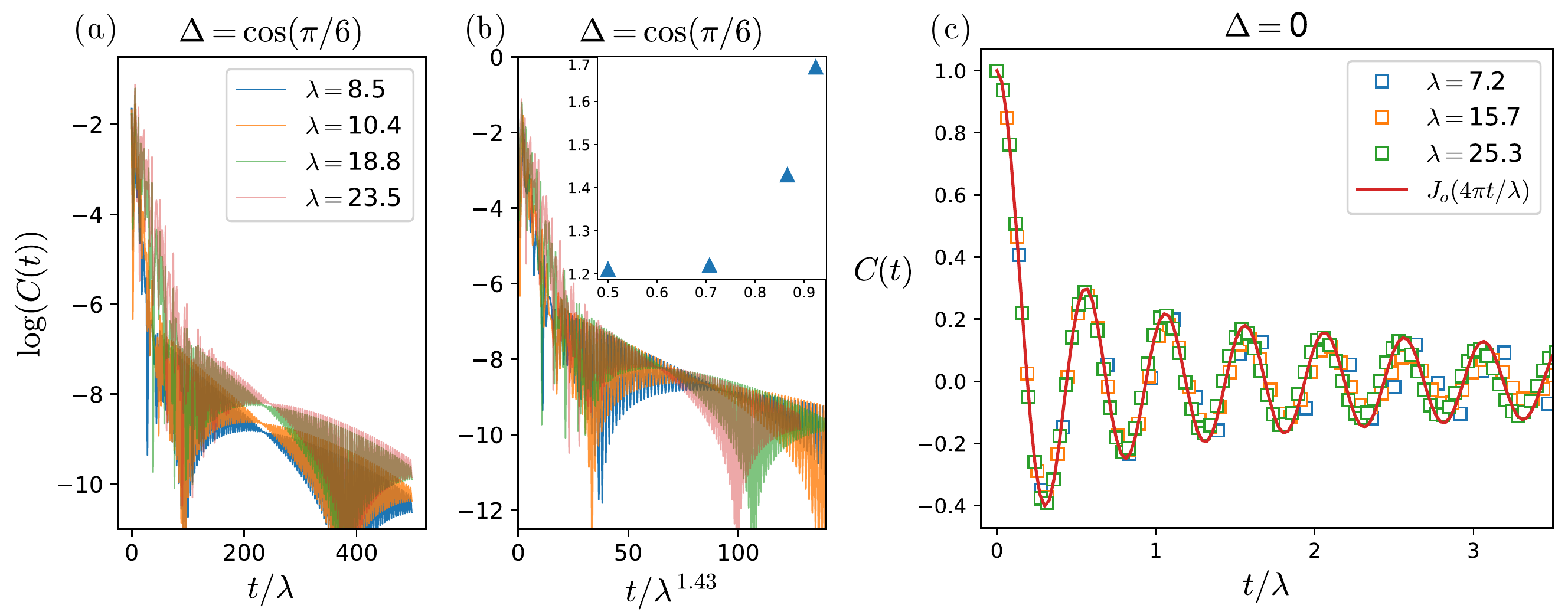}
  \caption{{\bf Easy-plane regime.} GHD prediction for the time evolution of the contrast \eqref{eq:contrast} from the helix initial state in the regime $0 \leq \Delta<1$. Left plots (a,b)  display the time evolution of the logarithm of the contrast under XXZ Hamiltonian with $\Delta=\cos \pi/6 \sim 0.87$, showing exponential decay at long times for any finite value of the helix wave-length $\lambda$. The ballistic time-dependence as $t/\lambda$ is violated in the displayed time-scales, whereas an approximate, $t/\lambda^{1.43}$ (superdiffusive) rescaling leads to an apparent collapse of the data for different $\lambda$ on the displayed time-scales. The inset of (b) displays the approximate exponent $\alpha$ in $t/\lambda^\alpha$ fitted for different values of $\Delta$ (see supplementary material for additional numerical data), showing an approach to diffusion $\alpha=z=2$ as $\Delta \to 1$. (c)  Time evolution of the contrast at the free fermion point $\Delta=0$, where the hydrodynamic prediction is written in a closed form in terms of a Bessel function, $C(t) = J_0(4 \pi t/\lambda)$, signalling exact ballistic dynamics. Exact numerical simulations at different values of  $\lambda$ are presented, showing very good agreement with the hydrodynamic predictions already for intermediate values of $\lambda$.      }\label{Fig:gapless}
\end{figure*}  
\paragraph{\textbf{Helices and local equilibrium} ---}
We consider the initial helix state 
\begin{equation}\label{eq:Helixstate}
| \psi_0 \rangle  = \bigotimes_{j=1}^L \left( \cos \frac{\theta_j}{2} \left| \uparrow \right.\rangle_j  + \sin \frac{\theta_j}{2} \left|\downarrow  \right.\rangle_j   \right),
\end{equation}
where the angle spins over a length given by $\lambda$ as $\theta_j = \theta_0 + (2 \pi j)/\lambda$. This state is characterized by an initial profile of magnetization $\langle S_j^z \rangle = 1/2 \cos \theta_j$ and energy $e_i = S^x_i S^x_{i+1} + S^y_i S^y_{i+1} + \Delta S^z_i S^z_{i+1}$, in the limit of large $\lambda$, given by 
\begin{equation}\label{eq:energydensity}
\langle e_i \rangle = \Big( 1+\Delta + (\Delta-1) \cos(\theta_j) \Big)/8. 
\end{equation}
We aim at computing the spin contrast,
\begin{equation}\label{eq:contrast}
C(t) = 4 \sum_j  \cos(\theta_j ) \langle \psi_0(t)| S^z_j | \psi_0(t) \rangle ,
\end{equation}
 which
quantifies the magnetization dynamics of the system and was computed experimentally~\cite{bloch_heisenberg,Jepsen2020}.  Let us first consider the case of an initial homogeneous state $| \psi(\theta_0) \rangle $ with $\theta_j=\theta_0$ for all $j$.   When letting such initial state time evolve under Hamiltonian \eqref{eqXXZ}, the system quickly reach a local equilibrium described by a Generalized Gibbs Ensemble (GGE) \cite{
Vidmar2016,PhysRevLett.109.247206,PhysRevLett.106.140405,PhysRevLett.115.157201}, namely for a generic local operator ${\mathcal O}$ at large (microscopic) times 
\begin{equation} \label{eqGGE}
\langle \psi_0 (\theta_0) | {\mathcal O}(t) | \psi_0(\theta_0) \rangle \to {\rm Tr}[\rho_{\rm GGE}(\theta_0)\  {\mathcal O} ],
\end{equation} 
where $\rho_{\rm GGE}(\theta_0)= e^{- \sum_i \beta^i(\theta_0) Q_i}/Z$, where $Q_i$ are all the conserved total operators of Hamiltonian \eqref{eqXXZ}, such that $[H,Q_i]=0$, and with the chemical potentials $\beta^i(\theta_0)$ depending non-trivially on the angle. While the helix state \eqref{eq:Helixstate} is a pure state, we expect that over a short time scale it will first thermalize locally $| \psi_0 \rangle \langle \psi_0 | \to \prod_i \rho_{\rm GGE}(\theta_i)$. This local equilibrium assumption corresponds to a local density approximation (LDA): the (pure) initial state is replaced by a local equilibrium state, which we expect to be valid for $\lambda \gg a$ with $a$ the lattice spacing.

\paragraph{\textbf{Hydrodynamics} ---}
 The resulting evolution at longer times from local to global equilibrium is then controlled by the theory of {\em hydrodynamics}. Notice that if the only conserved quantities of the system were energy and magnetization, we would need to evolve the initial energy and magnetization profiles (fixed from the initial state from LDA) using a 2-component hydrodynamic evolution. However, since the XXZ spin chain is integrable, the number of initial chemical potentials $\beta^i$ is infinite, and the correct hydrodynamic evolution is the recently introduced Generalized Hydrodynamics (GHD)~\cite{PhysRevLett.122.240606,PhysRevLett.123.130602,PhysRevX.6.041065,PhysRevLett.117.207201,SciPostPhys.2.2.014,PhysRevLett.125.240604,vir1,PhysRevB.101.180302,2005.13546,GV19,Gopalakrishnan2019,PhysRevB.102.115121,PhysRevB.96.081118,PhysRevLett.125.070601,PhysRevLett.124.210605,PhysRevLett.124.140603,Bulchandani_2019,SciPostPhys.3.6.039,Doyon_2017,PhysRevLett.128.190401,ruggiero2019quantum}. It works in the following way: the initial profile of chemical potentials can be recast into an initial profile of density of occupations of quasiparticles $\rho^{\theta_j}_s(u)$. Different species of quasiparticles are called strings and are labelled by the index $s$, and their momentum $k(u)$ and energy $\varepsilon(u)$ are parametrized by the rapidity parameter $u$. Their time evolution $\rho_s(u; x ,t )$ is given by the Navier-Stokes GHD equations, which read in full generality as~\cite{DeNardis2018}
\begin{equation}\label{eq:GHD}
\partial_t \rho_s + \partial_x (v^{\rm eff}_s \rho_s ) = \frac{1}{2} \partial_x \Big(\sum_{s'}\mathfrak{D}_{s,s'} \cdot \partial_x \rho_{s'} \Big),
\end{equation}
with initial condition $\rho_s(u; x ,t )=\rho^{\theta_{x=j}}_s(u)$. We stress that given the periodicity of the initial state, we can rescale space and time by $\lambda$, studying this way eq. \eqref{eq:GHD} in the system $x\in[0,1]$ with periodic boundary conditions and with the diffusive right hand side rescaled by $1/\lambda$. The effective velocity of the quasiparticles is defined as their group velocity $v^{\rm eff}_s(u) = \varepsilon'(u)/k'(u)$ (which depends non-trivially on the density), and the diffusion kernel $\mathfrak{D}_{s,s'}(u,u')$ gives the effective diffusion of each quasiparticles due to their local microscopic scatterings~\cite{DeNardis2018, DeNardis_SciPost, Gopalakrishnan18}. Equation \eqref{eq:GHD} is therefore strongly non-linear and it can be solved by simple generalizations of midpoint or backward Euler methods. 
The quasiparticle content is drastically different in the easy-axis regimes $|\Delta| \geq 1$ and the easy-plane.  We shall first consider the latter, which also includes the free fermionic point $\Delta=0$.  

\paragraph{\textbf{Easy plane regime} ---} We focus here on the regime $|\Delta| < 1$, see Fig. \ref{Fig:gapless}. Let us first consider the free fermions case $\Delta=0$, as also recently studied in \cite{PhysRevA.106.043306}. In this case, after a Jordan-Wigner transformation, the dynamics can be simulated exactly for any system size and time, as the initial state is a Gaussian state. We are therefore able to check exact numerical results against the hydrodynamic approximation, namely given by eq \eqref{eq:GHD} where $u=k$, $v^{\rm eff} = \sin (k)$ and zero diffusion. The hydrodynamic equations can in this case be solved exactly giving $\rho(k;x,t) = \rho^0_k(x- \sin (k) t)$, where $ \rho^0_k$ is obtained from the helix initial state using LDA~\cite{suppmat}. We find the contrast $C(t) = J_0(4 \pi t/\lambda)$ with $J_0$ a Bessel function, which decays to zero algebraically as $t^{-1/2}$, and not exponentially as assumed in Ref.~\cite{Jepsen2020}.
The comparison between the exact simulations, see Fig. \ref{Fig:gapless} and the hydrodynamic predictions shows a very good agreement even for relatively small $\lambda \sim O(10)$, confirming the validity of the hydrodynamic approach.

\begin{figure}[t]
  \includegraphics[width=0.45 \textwidth]{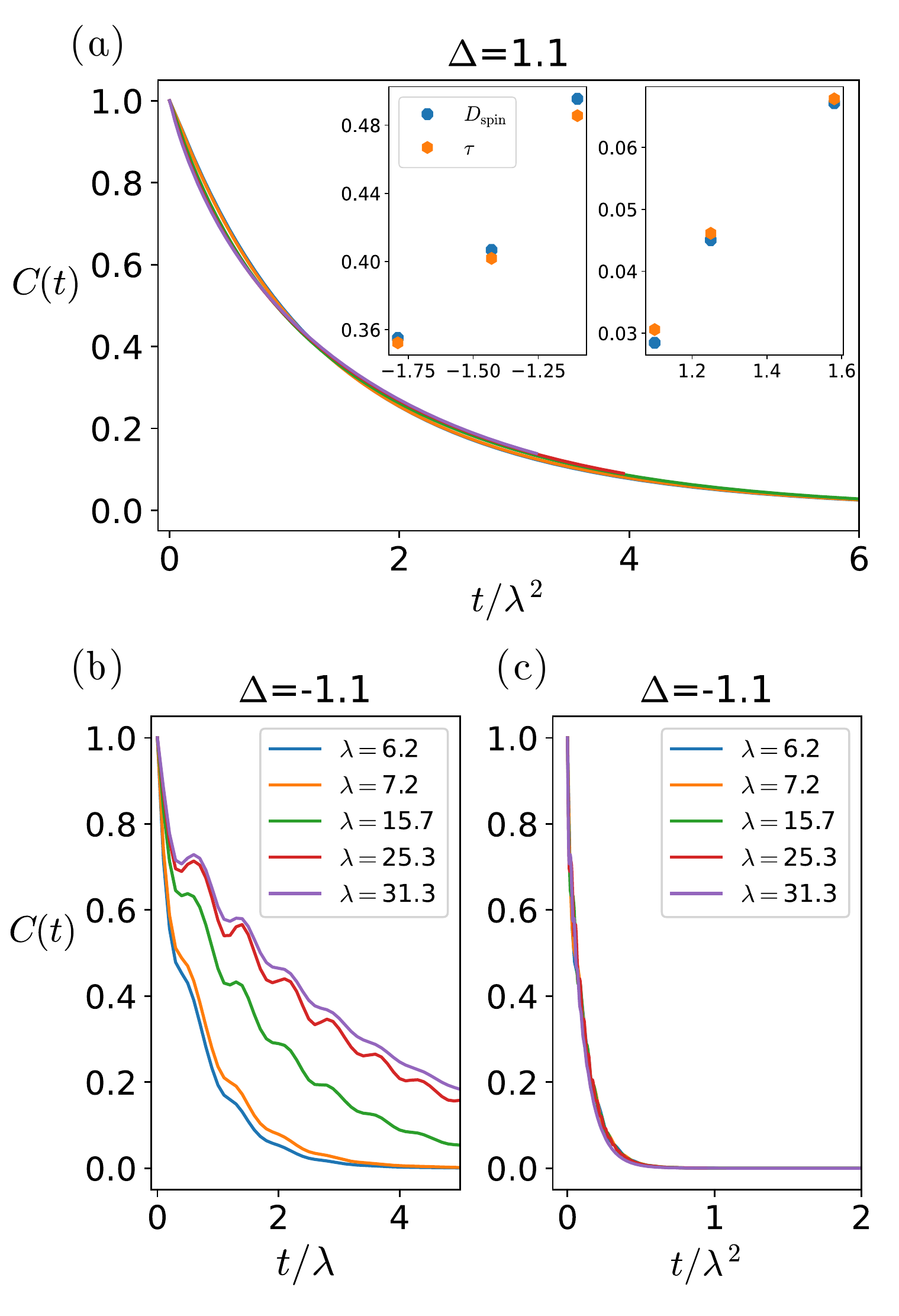}
  \caption{{\bf Easy-axis regime.}  GHD prediction for the time evolution of the contrast \eqref{eq:contrast} from the helix initial state in the regime $ |\Delta| >1$. Top plot (a)  displays the time evolution under XXZ Hamiltonian with $\Delta=1.1$, showing exponential decay with diffusive scaling $t/\lambda^{2}$ for the different values of $\lambda$ reported in plots (b,c). The inset displays the numerically extracted value of relaxation time $\tau = -  1/2 \lim_{t \to \infty}  (t/\lambda^2)^{-1} \log C(t)$ compared with the infinite time prediction of $D_{\rm spin}$ in eq. \eqref{eq:GHDSpin}, which becomes flat in $x$, see Fig. \ref{Fig:vD}.
   Plot (b) displays the same time evolution but with negative value of $\Delta=-1.1$, as function of $t/\lambda$, showing an, approximated, ballistic rescaling $t/\lambda$ at short times, with the long-time diffusive decay with $t/\lambda^2$ dependence showed in plot (c).      }\label{Fig:gapped}
\end{figure}  

We now turn to the interacting case, focusing on the values $\Delta = \cos (\pi /\ell)$ with $\ell > 2$ because the quasiparticle content is simpler at these points: for $\Delta = \cos(\pi/\ell)$ one has $\ell$ quasiparticle species, each of them with their associated $\rho_s(u)$~\cite{PhysRevB.96.020403,PhysRevB.101.224415,PhysRevLett.82.1764}, and the magnetisation profile at time $t$ is given by 
\begin{equation}
\langle S^z(x,t) \rangle = 1/2 - \sum_{s=1}^\ell  s \int du \  \rho_s(u; x,t).
\end{equation}
The GGE corresponding to the initial state can be found using the transfer-matrix based approach introduced in \cite{PhysRevB.96.020403}.
Relative to $\Delta = 0$, the key new feature of the interacting case is that the ballistic propagation of quasiparticles is convolved with diffusive spreading due to elastic collisions~\cite{DeNardis2018, DeNardis_SciPost, Gopalakrishnan18}.
For an initial state of fixed $\lambda$, this convolution transforms into a product, and the contrast goes as $C(\lambda,t) = f(t/\lambda) \exp(- D t/(2\lambda^2))$, where $D$ is an effective spin diffusion constant. 
In the Euler scaling limit $t \to \infty, \lambda \sim t$, the diffusive correction becomes irrelevant; however, the late-time limit for fixed $\lambda$ is dominated by this diffusive correction. 
If we fit the data over intermediate time ranges, we find an apparent superdiffusive collapse $C(t) = g(t /\lambda^\alpha)$ with $1 < \alpha < 2$, consistent with the experiment, though our analysis suggests that this is a finite-time effect (and indeed $\alpha$ drifts toward $2$ as we fit later times). 
Moreover, as $\ell \to \infty$ (i.e., $\Delta \to 1^-$), this apparent $\alpha$ drifts toward $2$, because $D$ diverges in this limit and the contrast is increasingly dominated by the diffusive correction: in order to recover the ballistic scaling one would need to go to inaccessibly late times and long wavelengths. We will return below to the limiting behavior at $\Delta = 1$.

 %The reason for such a choice of anisotropy relies entirely on the simpler structure of the quasiparticle content. 
%
%In this case indeed the model is described by $\ell$ quasiparticles, each of them with their associated $\rho_s(u)$, see Refs.~\cite{PhysRevB.96.020403,PhysRevB.101.224415,PhysRevLett.82.1764}. The initial profiles of GGE can be found using the transfer-matrix based approach introduced in \cite{PhysRevB.96.020403}. We find results that are compatible with a long time ballistic scaling $C(t) = f(t/\lambda)$, with $t,\lambda \to \infty$ and $t/\lambda$ fixed, in agreement with linear-response expectations. However, for fixed $\lambda$, our data indicates exponential decay at long times due to the diffusive corrections in eq.~\eqref{eq:GHD}, with the contrast given by $C(t) = {\rm e}^{-D t/\lambda^2 } f(t/\lambda)$. While the asymptotic scaling (for $t \to \infty$, $t \sim \lambda$) is ballistic, at fixed $\lambda$ and long times, the exponential decay dominates, while for intermediate times we find that we can collapse the data using a superdiffusive scaling $C(t) = g(t /\lambda^\alpha)$, with $1<\alpha<2$, approaching $\alpha \to 2$ near the isotropic limit. This apparent {\em superdiffusive} scaling was also observed experimentally~\cite{Jepsen2020}, though we caution that it likely results from finite-time and $\lambda$ effects.

\paragraph{\textbf{Easy axis regime} ---} The regime $|\Delta| \geq 1$ is characterized by an infinite number of quasiparticles, normalized such that their integrated sum gives the value of the absolute value of the magnetization $|\langle S^z \rangle | = 1/2 - \sum_s s \int \ du \rho_s(u)$. As quasiparticles can only provide the absolute value of magnetization it is clear that the full magnetization field requires some extra information, namely the sign of the magnetization. As first introduced in \cite{piroli2017,dbd2}, the positions of the domain walls where the sign of the magnetization changes is given by the positions $ x_\infty(t)$ of the largest quasiparticles, with label $s \to \infty$, the so-called giant magnons, which move with effective velocity $v_{\rm spin}= v^{\rm eff}_{\infty}$ and which need to be treated aside. The inclusion of diffusion gives a Gaussian spreading to the position of the largest quasiparticles, given by the diagonal element of the diffusion kernel ${D}_{\rm spin}  = \lim_{s \to \infty} \mathfrak{D}_{s,s} $,  which can then be evolved via the equation
\begin{equation}\label{eq:GHDSpin}
\partial_t \mathfrak{f} + v_{\rm spin} \partial_x  \mathfrak{f} = \frac{1}{2} \partial_x \Big( {D}_{\rm spin} \partial_x \mathfrak{f} \Big),
\end{equation}
with initial condition $ \mathfrak{f} = {\rm sgn}(\langle S^z(x,t) \rangle)$. 
Then magnetization is obtained by evolving this together with \eqref{eq:GHD} and computing $\langle S^z(x,t) \rangle = \mathfrak{f}(x,t) [1/2 - \sum_s  s \int \ du \rho_s(u;x,t)]$. We evolve from the initial GGE fields, using the expression for the densities already being found explicitly in \cite{PhysRevB.94.054313}.

Let us first focus on the regime $\Delta = 1+ \epsilon^+$. In this case the initial helix state is locally close to the ferromagnetic vacuum, where energy density equal to $1/4$.  Such thermodynamic states are characterised by vanishing spin velocity and very small spin diffusion constant, which eventually goes to zero at $\Delta=1$. For any $\Delta>1$ the dynamics is diffusive with a relativity small diffusion constant,  which we believe is responsible for apparent finite-time subdiffusive scaling observed experimentally in that regime~\cite{Jepsen2020}, see Fig. \ref{Fig:gapped}. Note that our hydrodynamic equations do not include subdiffusive corrections, but they predict asymptotically diffusive relaxation, albeit with a minute diffusion constant.

\begin{figure}[t]
  \includegraphics[width=0.49 \textwidth]{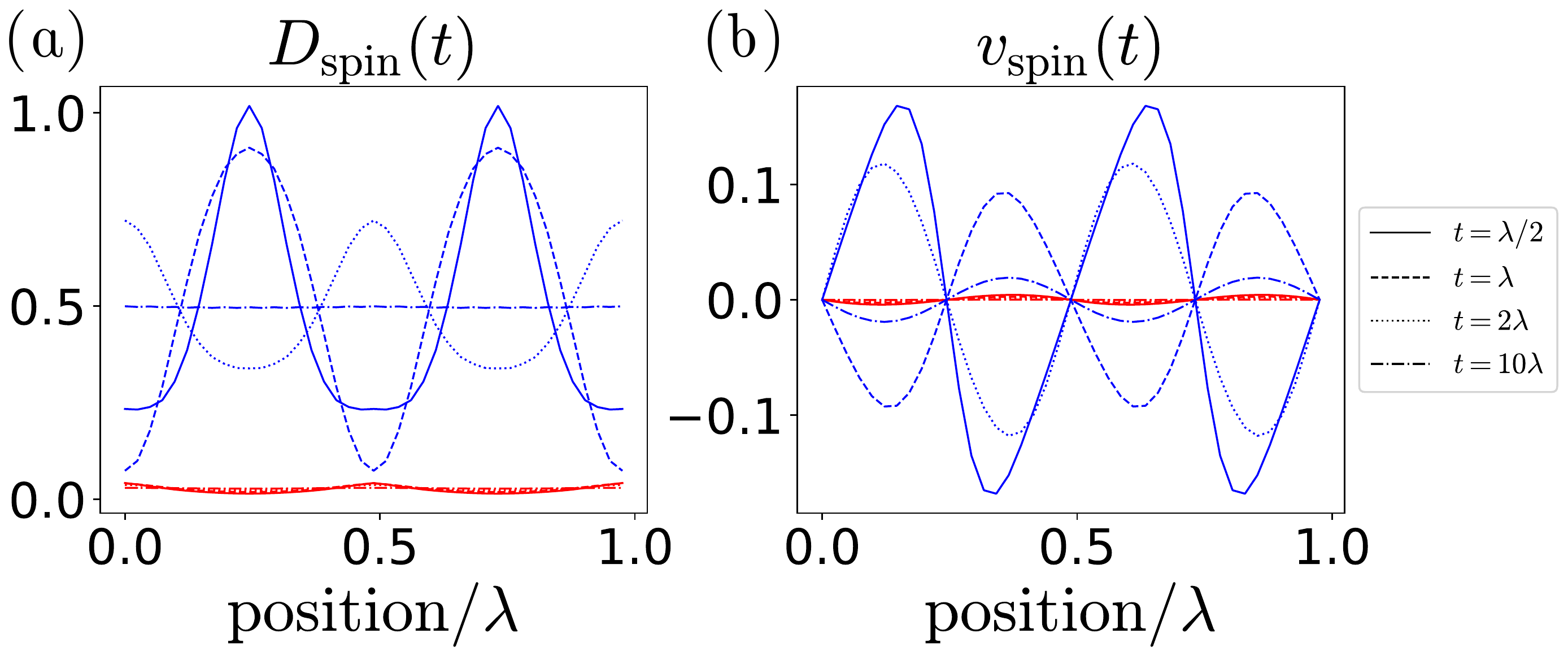}
  \caption{{\bf Short time effects.} Time evolution of spin velocity (a) and diffusion constant (b) in eq. \eqref{eq:GHDSpin}, plotted as function of position $j/\lambda$ at different times and for two different values of $\Delta$, one positive $\Delta=1.1$ (red lines) and one negative $\Delta=-1.1$ (blue lines). We show how for positive $\Delta$, velocity and diffusion constant remains constantly very small (consistent with apparent subdiffusion at short times). For negative $\Delta$, the diffusion constant is finite and converges at large times to a constant function in $x$, while the spin velocity also grows in time (from its initial condition $v_{\rm spin}(x,t=0)=0$), to then later decay again to zero, explaining short-time ballistic dynamics. } \label{Fig:vD}
\end{figure}

The regime $\Delta < - 1$ presents some surprises, see Fig. \ref{Fig:gapped}. From linear response results, we expect that no spin ballistic transport should be present at zero net magnetization in the system. However in this regime we witness short-time ballistic dynamics given by an effective thermoelectric effect. The initial energy density is not flat, contrary to the case $\Delta \sim 1^+$, see \eqref{eq:energydensity}. This is true for most of the initial densities of conserved quantities but we can restrict ourself to energy density in order to explain the main physical effects.  Such an initial unbalance is initially ballistically redistributed (given their finite velocities $v^{\rm eff}_s(u)$) by the small $s \sim O(1)$ quasiparticles in the system, which are spin uncharged. Such a flow of small quasiparticles in the system induced a finite velocity for the largest quasiparticle, $v_{\rm spin}$, which is pushed by chiral scatterings with the lighter ones, see Fig. \ref{Fig:vD}. At short times we thus see signatures of ballistic transport, although ultimately the contrast decays diffusively as $C(t) \sim {\rm e}^{- D
 t/(2\lambda^2)}$ with a prefactor $D$ given by the spin diffusion constant of the global equilibrium final state, see Fig. \ref{Fig:vD}. This short time ballistic dynamics for $\Delta <-1$ was also observed experimentally~\cite{Jepsen2020} and had remained unexplained until now: note that this result is particularly surprising from the point of view of linear response since high temperature transport does not depend on the sign of $\Delta$. This asymmetry between $\Delta>1$ and $\Delta<-1$ regimes relies entirely on the special nature of the helix initial state and on thermoelectric effects. 

\paragraph{\textbf{Discussion and the point $\Delta=1$} ---}  We have developed a hydrodynamic description of the relaxation of spin spirals in the XXZ spin chain, based on imposing local equilibrium using the local density approximation. This description captures all the experimentally observed relaxation phenomena, with one important exception: the case of $\Delta = 1$. Here, the local density approximation incorrectly predicts that a long-wavelength spin spiral does not relax at all, since it is locally in the quasiparticle vacuum. In fact, experiments see relaxation with dynamical exponent $z = 2$~\cite{bloch_heisenberg}. To describe this case, one must go beyond the local density approximation. We briefly outline how the $z = 2$ scaling follows from GHD. We imagine cutting the system up into hydrodynamic cells on a much larger length-scale than $\lambda$, and assuming that each cell equilibrates. Because the initial condition is smoothly modulated at $\lambda$, its quasiparticle content will be dominated by strings of size $s \sim \lambda$ (this is the crucial distinction between the helix and a thermal state, which has a string population involving all sizes).  Relaxation occurs when these dominant strings cross a distance $\lambda$. Since at $\Delta=1$, a $s=\lambda$-string has in general a velocity $v^{\rm eff}_s \sim 1/s = 1/\lambda$ \cite{PhysRevX.11.031023}, at large $\lambda$, the associated timescale scales as $\lambda^2$, yielding $z = 2$. Incorporating this physics more quantitatively in our framework remains a task for future work.

More generally, our results suggest that GHD (supplemented with diffusive corrections) remains a powerful framework for describing the dynamics of initial states that are far from local thermal equilibrium. It would be interesting to extend our framework to other far-from-equilibrium states---e.g., those following interaction quenches or the Newton's cradle setup~\cite{kinoshita2006quantum, le2022direct}---and to incorporate experimental features such as trap-induced inhomogeneity and other integrability-breaking perturbations.

{\it Acknowledgements.}—We thank Vir Bulchandani, Wen Wei Ho, Andrea De Luca, and Michael Knap for helpful discussions. This work was supported by the ERC Starting Grant 101042293 (HEPIQ) (J.D.N.), the National Science Foundation under NSF Grant No. DMR-1653271 (S.G.),  the US Department of Energy, Office of Science, Basic Energy Sciences, under Early Career Award No. DE-SC0019168 (R.V.), and the Alfred P. Sloan Foundation through a Sloan Research Fellowship (R.V.). 

\bibliography{SSD,refs,biblio}

\onecolumngrid
%\appendix

\newpage 

\begin{center}
\textbf{\large Supplementary Material} \\ 
\end{center}
\label{appmain}

\section{Contrast at $\Delta=0$}

The Hamiltonian \eqref{eqXXZ} with $\Delta=0$ can be mapped via Jordan-Wigner transformation into the tight binding model 
\begin{equation}
H =  \sum_j  c^\dagger_j c_{j+1} + h.c.
\end{equation}
with fermionic operator $\{ c_i , c^\dagger_j \} = \delta_{ij}$. 
The initial helix state \eqref{eq:Helixstate} is a Gaussian state and its correlation matrix can be computed exactly 
\begin{equation}
C_{ij} = \langle \psi_0 | c^\dagger_i c_j | \psi_0 \rangle = \delta_{ij}\cos^{2}(\theta_{j}/2)+(1-\delta_{ij})\sin^{2}(\theta_{i}/2)\cos^{2}(\theta_{j}/2),
\end{equation}
which can be evolved with the single-particle Hamiltonian to obtain exact numerical simulations. The hydrodynamic limit can be obtained by computing the momentum occupation function $n(k) = \langle c^\dagger_k c_k \rangle = 2 \pi \rho(k)$ at each point $x=i$, and then its GHD time evolution 
\begin{equation}
\partial_t n(k;x,t) + v_k \partial_x   n(k;x,t)=0,
\end{equation}
is simply solved by $n(k;x,t) = n(k;x- v(k) t)$, with $v(k) = \sin k$,  which reads 
\begin{equation}
n(k; x , t )=  \cos^4 \left(\frac{\theta(x-v_kt)}{2} \right)+ 2\pi\delta(k)\sin^2 \left(\frac{\theta(x-v_kt )}{2} \right) \cos^2 \left(\frac{\theta(x-v_kt)}{2} \right) ,
\end{equation}
with $\theta(x) = 2 \pi x/\lambda$.  The contrast is then given by
\begin{align*}
    C(t)&= \int_{0}^{\lambda} dx \left[ \frac{1}{2} -\int_{-\pi}^{\pi} \frac{dk}{2 \pi} n(k; x-v(k)t) \right] \sin \left(\frac{2 \pi x}{\lambda} \right)   = \frac{1}{2\pi} \int_{-\pi}^{\pi} dk \; \frac{\lambda}{4}\cos\left(\frac{\pi}{2}+\frac{4 \pi t}{\lambda} \sin(k) \right) = J_0(4 \pi t /\lambda).
\end{align*}

\section{Additional numerical data in the easy plane regime}

\begin{figure*}[h!]
  \includegraphics[width=\textwidth]{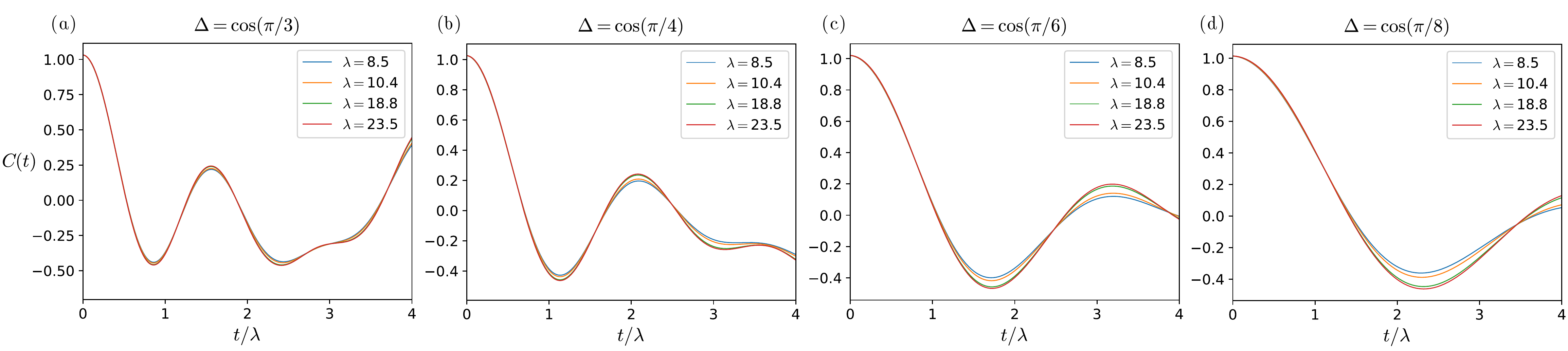}
  \caption{{\bf Easy-plane regime.} GHD prediction for the time evolution of the contrast \eqref{eq:contrast} from the helix initial state in the regime $0 \leq \Delta<1$ at short times, plotted as function of $t/\lambda$.  We see how upon increasing $\Delta$ the dynamics stop being purely ballistic at shorter times.    }\label{Fig:gapless-1}
\end{figure*}

\begin{figure*}[h!]
  \includegraphics[width=0.5\textwidth]{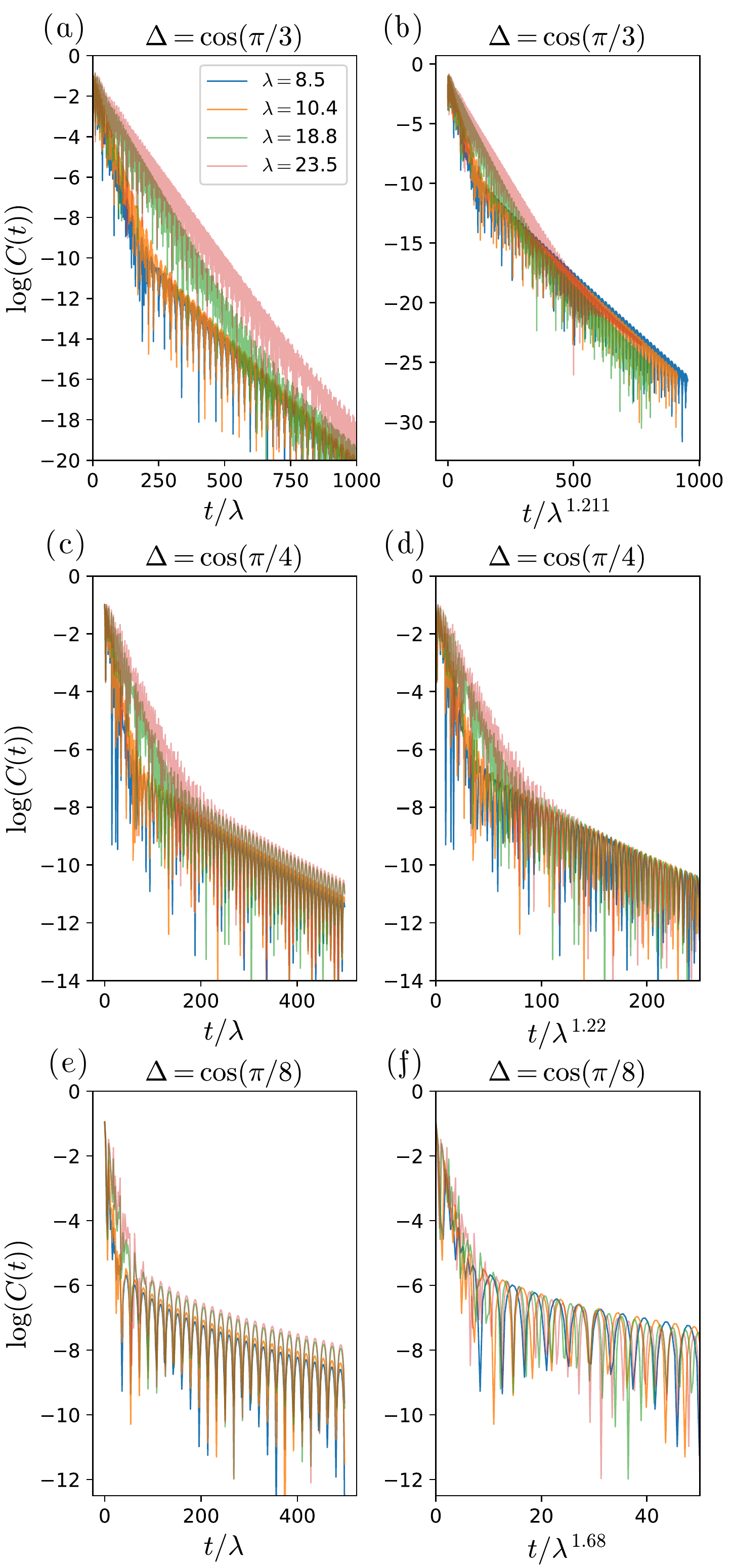}
  \caption{{\bf Easy-plane regime.} GHD prediction for the time evolution of the contrast \eqref{eq:contrast} from the helix initial state in the regime $0 \leq \Delta<1$ at short times, plotted as function of $t/\lambda$ (left) and of the rescaled (superdiffusive) time $t/\lambda^\alpha$, using the same values reported in Fig. \ref{Fig:gapless} in the main text, to achieve collapse of the data.     }\label{Fig:gapless-2}
\end{figure*}  

\end{document}